\documentstyle[12pt]{article}
\begin{document}
\textheight 7.6in
\begin{flushright}
\mbox{BA-99-09}\\
%\mbox{hep--ph/99}\\
\mbox{January 1999} \\[0.5in]
\end{flushright}
\begin{center}
{\Large\bf Relaxing Axions}\\[0.5in]
{\large\bf S.M. Barr$^1$ \\
Bartol Research Institute \\ University of Delaware \\
Newark, DE 19716}\\[0.4in]
\end{center}

%\date{}
%\maketitle

\begin{abstract}

A mechanism for lifting the cosmological upper bound on the axion decay
constant, $f_a$, is proposed. It entails the near masslessness of the 
radial mode whose vacuum expectation value is $f_a$. Energy in the
coherent oscillations of the axion field in the early universe gets fed
into the motion of the radial mode, from which it is then redshifted
away. It is found that the initial value of $f_a$ can be at scales
between $2 \times 10^{14}$ GeV and $10^{16}$ GeV. This evolves with
time to values close to the Planck scale. It is suggested
that the nearly massless radial mode might play the role of
quintessence.
\\[0.3in]
\end{abstract}
$\line(75,0){75}$\\
\noindent
$^1$E-mail: smbarr@bartol.udel.edu \\

\newpage

\section{Introduction}

The axion$^1$ idea is an extremely elegant approach to solving the Strong 
CP Problem. However, it is not without difficulties. The main difficulty
is that axions have neither been observed directly in the laboratory,
nor indirectly through astrophysical effects. This implies that the axion
decay constant, $f_a$, must be greater than about $10^{10}$ GeV.
Moreover, there is the cosmological bound, based on the so-called
``axion energy problem",$^2$ that $f_a$ is less than about $10^{12}$ GeV.
This means that the axion decay constant (or, equivalently, the scale
at which the Peccei-Quinn symmetry is broken) cannot be at any of the
mass scales at which other physics is known or suspected to be based.
In particular, it cannot be at the weak scale, the Planck scale, or
the grand unification scale. 

A number of attempts$^3$ have been made to weaken or remove the 
cosmological bound on $f_a$. In this paper we suggest an idea for
doing this. The idea is that if the radial mode associated
with the axion (that is, the scalar field whose expectation value 
is $f_a$) has a nearly flat potential then it would have increased
with time as soon as the coherent axion oscillations commenced.
As a result of this, energy would have drained out of the axion oscillations
into the nearly massless radial mode, and from thence been redshifted 
away by the cosmic expansion.
We find that the initial value of $f_a$ can be as large as
$10^{16}$ GeV. In the course of cosmic evolution this would have
``relaxed" to a value we presume to be near the Planck scale. Axions 
in this scenario would therefore be extremely hard if not impossible 
to detect. However, the radial mode may play the role of ``quintessence".$^4$ 

The plan of the paper is as follows. In section 2, a rough sketch
of the axion energy problem in conventional axion models is given. 
In section 3, a simplified discussion of the relaxing axion
scenario is given to show how energy is drained out of the axion 
field. In section 4, a more detailed quantitative analysis is given, 
still assuming, however, that the radial direction is exactly flat. 
In section 5, the radial potential is discussed, in particular
how flat it must be and how such flatness might arise. 

\section{The usual axion energy problem}

In ordinary axion models, the Peccei-Quinn symmetry is spontaneously
broken by a complex scalar field, which we shall denote $\Phi$. The
expectation value of $\Phi$ is called $f_a$, while the phase
of $\Phi$ is the axion mode. Thus one may write $\Phi = r e^{i \theta}$, 
where $\langle r \rangle = f_a$, and the axion field $a$ is given by 
$a = f_a \theta$. Since the Peccei-Quinn symmetry is
anomalous, QCD instanton effects give the axion a mass. 
The instanton-generated potential has the form $V_I =
- \mu_0^4 \cos (N \theta)$, where $\mu_0$ is of order the QCD scale, and 
$N$ is an integer, which for present purposes can be taken to be 1.

When the temperature in the early universe was well above the QCD 
scale the instanton-generated potential
was not fully turned on, whereas after the temperature fell below 
the QCD scale it had essentially its full zero-temperature strength.
Let us call the time when the instanton potential reached
nearly full strength, $t_0$. Since the temperature was then of order 
$\mu_0$, $t_0 \sim M_P/\mu_0^2$. Assuming a prior epoch of inflation,
the axion field at $t_0$ was approximately spatially constant, but
barring an extremely unlikely coincidence, it had no reason to
be sitting at the minimum of its potential (which we take to be
at $\theta = 0$). Rather, it had some arbitrary value $\theta ( t_0) =
\theta_0$. Consequently, the axion field underwent coherent oscillations 
about its minimum. The energy density in these oscillations at time 
$t_0$ would have been approximately $\frac{1}{2} \mu_0^4 \theta_0^2$.
This energy scaled as $R^{-3}$, where $R$ is the scale factor of
the universe, and therefore the energy density
in the coherent axion oscillations remained (after $t_0$)
in a constant ratio to the baryon energy density. Since at $t_0$ the 
baryon energy density was of order
$10^{-10} m_p \mu_0^3$, the present ratio of
axion to baryon energy is roughly $(\frac{1}{2} \theta_0^2 \mu_0^4)/
(10^{-10} m_p \mu_0^3) \sim 10^9 \theta_0^2$. If the axions are not to
overclose the universe, $\theta_0^2 \stackrel{_<}{_\sim} 10^{-7}$.

This bound on $\theta_0$ can be satisfied if $f_a$ is sufficiently small.
The point is that the instanton-generated potential of the axion
field did not turn on instantaneously. If the potential turned on
slowly then the number of quanta in the coherent axion oscillations
remained constant since it is an adiabatic invariant. This implies that as
the mass of the axion increased the amplitude of the oscillations
decreased, and therefore by $t_0$, when the potential reached
full strength, the amplitude $\theta_0$ could have been very small. A simple 
estimate gives that $\theta_0^2 \sim f_a/M_P$. Thus, to satisfy the bound 
on $\rho_{axion}$ one requires that $f_a \stackrel{_<}{_\sim} 10^{12}$ GeV.

\section{The relaxing axion mechanism}

Let us suppose that the field $\Phi$ whose phase is the axion
has a flat potential in the radial direction. Then the
Lagrangian density can be written

\begin{equation}
\begin{array}{lcl}

L & = & \frac{1}{2} \left| \partial_{\mu} \Phi \right|^2 
+ \mu_0^4 \cos \theta, \\
& & \\
& \cong & \frac{1}{2}(\partial_{\mu} r)^2 
+ \frac{1}{2} r^2 (\partial_{\mu} \theta)^2 - \frac{1}{2} \mu_0^4 \theta^2.

\end{array}
\end{equation}

\noindent
As before, $\Phi = r e^{i \theta}$, and the axion field is $a = r \theta$.
Ignoring spatial derivatives of the fields, the equations of motion of 
$r$ and $\theta$ are

\begin{equation}
\begin{array}{lcl}

0 & = & \ddot{r} + 3 H \dot{r} - r \dot{\theta}^2, \\
& & \\
0 & = & \ddot{\theta} + (3 H + 2 \frac{\dot{r}}{r}) 
\dot{\theta} + \mu_0^4 r^{-2} \theta.

\end{array}
\end{equation}

\noindent
The last term in the equation for the radial mode $r$ is just
the centrifugal force, and it is this that drives $r$ to larger
values as $\theta$ oscillates.

The energy densities in the radial and axion modes are given simply
by $\rho_r = \frac{1}{2} \dot{r}^2$ and $\rho_{\theta} =
\frac{1}{2} r^2 \dot{\theta}^2 + \frac{1}{2} \mu_0^4 \theta^2$.
The energies in a comoving volume in these modes are given by
$E_r = R^3 \rho_r$ and $E_{\theta} = R^3 \rho_{\theta}$. Using the
equation of motion of $\theta$ to eliminate $\ddot{\theta}$, it
is easy to show that 

\begin{equation}
\dot{E}_{\theta} = R^3 \left[ - \frac{\dot{r}}{r} r^2 \dot{\theta}^2
+ 3 H (- \frac{1}{2} r^2 \dot{\theta}^2 + \frac{1}{2}
\mu_0^4 \theta^2) \right].
\end{equation}

\noindent
If the oscillator parameters evolve adiabatically, then averaged over
many oscillations the kinetic and potential energy in the oscillator
should be equal. That is, $\langle \frac{1}{2} r^2 \dot{\theta}^2 \rangle
= \langle \frac{1}{2} \mu_0^4 \theta^2 \rangle = \frac{1}{2} \langle 
\rho_{\theta} \rangle$. Therefore, averaged over many oscillations,

\begin{equation}
\dot{E}_{\theta} = - \frac{\dot{r}}{r} E_{\theta}
\end{equation}

\noindent
In a similar way, using the equation of motion of $r$ to eliminate
$\ddot{r}$, and averaging over many oscillations, one finds that

\begin{equation}
\dot{E}_r = \frac{\dot{r}}{r} E_{\theta} - 3 H E_r.
\end{equation}

\noindent
The interpretation is clear. The increase in $r$ caused by the centrifugal
force drains energy out of the axion oscillations and into the radial
mode, while at the same time energy is redshifted away from the 
radial mode. To put it another way, as $r$ increases the effective
mass of the axions, $\mu_0^2/r$, decreases, while the number of axions
remains constant. Thus $E_{\theta} \sim 1/r$, as implied also by
Eq. 4.

Suppose that $r \sim t^q$, and  
$E_{\theta} \sim t^{-q}$. The exponent $q$ can be determined by writing
the equation of motion of $r$ in terms of $E_{\theta}$ as follows:
$0 = \ddot{r} + 3 H \dot{r} - R^{-3} \langle E_{\theta} \rangle
r^{-1}$.
The first two terms go as $t^{(q-2)}$, while the last goes as 
$t^{-(2q+3/2)}$, assuming that $R \sim t^{1/2}$. Therefore $q = 1/6$.
Writing $E_{\theta} = E_{\theta 0} (t/t_0)^{-1/6}$, Eq. 5 is 
solved by $E_r = \frac{1}{8} E_{\theta 0} (t/t_0)^{-1/6}
+ c t^{-3/2}$. Thus, for large $t$, the radial energy is one-eighth
of the energy in the coherent axion oscillations, and the energy in
a comoving volume in either mode falls off as $t^{-1/6}$. 

\section{A more detailed analysis}

So far we have not taken into account the temperature dependence
of the instanton-generated potential of the axion field.
According to the dilute-instanton-gas calculation of Gross, Pisarski,
and Yaffe,$^5$ the instanton potential for $\theta$ at high temperature goes
as $\mu^4 \sim T^4$ $\exp (- 8 \pi^2/
g^2(T)) \sim T^4$ $\exp (\frac{1}{3} (11 N - 2 N_f) \ln T) \sim$
$T^{(-7 + 2 N_f/3)}$. We shall assume henceforth that $\mu^4 
= \mu_0^4 (T_0/T)^{2k} = \mu_0^4 (t/t_0)^k$, for $T \gg \mu_0$
(i.e. $t \ll t_0$), and that
$\mu^4 \cong \mu_0^4$ for $t \gg t_0$.
The dilute-instanton-gas calculation suggests that $k \cong 5/2$, but 
we shall keep $k$ as a parameter.

If we change variables to $\tau \equiv t/t_0$, and denote
$\partial/\partial \tau$ by a dot, then we can write the equations of motion

\begin{equation}
\begin{array}{lcl}
0 & = & \ddot{r} + \frac{3}{2 \tau} \dot{r} - r \dot{\theta}^2, \\
& & \\
0 & = & \ddot{\theta} +(\frac{3}{2 \tau} + 2 \frac{\dot{r}}{r}) \dot{\theta}
+ (\mu_0^4 t_0^2/r^2) \tau^k \theta. \\
& & 
\end{array}
\end{equation}

\noindent
The effective mass of the field $\theta$, then, is 
$m_{\theta} (\tau) = (\mu_0^2 t_0/r(\tau)) \tau^{k/2}$.
Therefore it is reasonable to make the ansatz that 
$\theta(\tau)$ has the form

\begin{equation}
\theta(\tau) = A(\tau) e^{i B(\tau)},
\end{equation}

\noindent
where $A$ and $B$ are real and 

\begin{equation}
\dot{B}(\tau) = m_{\theta}(\tau) = (\mu_0^2 t_0/r(\tau)) \tau^{k/2}.
\end{equation}

\noindent
Substituting into the equation of motion for $\theta$ and taking the
real and imaginary parts of the equation, one gets (using the fact
that $\dot{r}/r = k/2\tau - \ddot{B}/\dot{B}$)

\begin{equation}
\begin{array}{lcl}
0 & = & \ddot{A} +
\left[ (\frac{3}{2} + k)/\tau - 2 \ddot{B}/\dot{B} \right] \dot{A}, \\
& & \\
0 & = & -\ddot{B} + \left[ (\frac{3}{2} + k)/\tau + 2\dot{A}/A \right]
\dot{B}. \\
& &
\end{array}
\end{equation}

\noindent
The first of these equations is exactly solved by
$(\dot{A}/\dot{B}^2) \tau^{(3/2 + k)} =$ constant, while the second
is solved exactly by $(\dot{B}/A^2) \tau^{-(3/2 + k)} =$ constant. 
Eliminating $\dot{B}$ from these equations and solving gives

\begin{equation}
\begin{array}{lcl}
A & = &\theta_0 (\tau^{(5/2 +k)} + C))^{-1/3}, \\
& & \\
\dot{B} & = & c' \theta_0^2 (\tau^{(5/2 + k)} + C)^{-2/3} \tau^{(3/2 + k)}, \\
& & 
\end{array}
\end{equation}

\noindent
where $c'$, $C$, and $\theta_0$ are integration constants.
After many oscillations the integration constant $C$ can be neglected,
and one has, using Eqs. 7 and 10, a solution for $\theta(\tau)$
of the following form:

\begin{equation}
\theta(\tau) = \theta_0 \tau^{-(5/6 + k/3)} \exp (i b_0 \tau^{(5/6 + k/3)}
+ i b'_0).
\end{equation}

\noindent
($b_0 = \frac{c' \theta_0^2}{5/6 + k/3}$.) 
Eqs. 8 and 10 directly give the solution for the radial mode:
$r(\tau) = \frac{\mu_0^2 t_0}{(5/6 + k/3) b_0} \tau^{(1/6 + k/6)}$
(again, neglecting the integration constant $C$).
Note that for the case of $k = 0$, which corresponds to a fixed 
value of $\mu^4$, the radial variable increases as $\tau^{1/6}$,
in agreement with the result obtained in section 3.

It must be checked that this solution for $r(\tau)$
and the solution for $\theta(\tau)$ given in Eq. 11 
satisfy the equation of motion of $r$. Assuming that $\dot{\theta}$ 
is dominated by the rapid oscillations of $\theta$ rather than by the
slow variation of its amplitude, one has that $\dot{\theta}
\cong i b_0 (5/6 + k/3) \theta_0 \tau^{-1} \exp (i b_0 \tau^{(5/6 + k/3)}
+ i b'_0)$. Averaged over many oscillations, therefore,
$\langle \dot{\theta}^2 \rangle = \frac{1}{2} b_0^2 (5/6 + k/3)^2 \theta_0^2
\tau^{-2}$. Substituting this and the expression for $r(\tau)$ into 
the equation of motion for $r$ (see Eq. 6), one sees that that
equation is satisfied if 
$b_0 = \frac{ \sqrt{(1 + k)(4 + k)}}{5 + 2k} \theta_0^{-1}$.
Thus we have that

\begin{equation}
r(\tau) = \frac{6 \mu_0^2 t_0}{\sqrt{(1 + k)(4 + k)}} \theta_0 
\tau^{(1/6 + k/6)}.
\end{equation}

The solutions given in Eqs. 11 and 12 apply to the period $\tau < 1$,
in which the instanton potential was still turning on.
The same expressions with $k$ set to zero apply to the period 
$\tau > 1$, after the instanton potential turned on. 
The true solution will smoothly interpolate between these 
in the period $\tau \sim 1$. (Note that the $k \neq 0$ and $k = 0$
solutions for $\theta$ actually agree at $\tau = 1$, while the $k \neq 0$ and
$k = 0$ solutions for $r$
differ at $\tau = 1$ by a factor of $\sqrt{(1 + k)(4 + k)}/2
\approx 2.4$.)

One is now able to estimate the energy in the axion oscillations.
The crucial parameter is the value of $r$ at the time when
the axion oscillations started. We will call that time $t_i$.
The oscillations started when the effective mass of the $\theta$
field, $m_{\theta} = \dot{B}$, became equal to the expansion
rate of the universe, $H = (2t)^{-1}$. That is,
when $(2 t_i)^{-1} \approx (\mu_0^2 t_0/r(t_i)) (t_i/t_0)^{k/2}$.
Clearly, the smaller $r(t_i)$ was, the earlier the axion oscillations began.
Turning this around, $r(t_i) \approx 2 \mu_0^2 t_0 (t_i/t_0)^{(1 + k/2)}
\sim M_P (t_i/t_0)^{(1 + k/2)}$.

At $t_i$ it is to be expected that $\theta$ was of order unity,
since it had not had time to be affected by the instanton potential.
But according to Eq. 11, $\theta(t_i) \sim \theta_0 
(t_i/t_0)^{-(5/6 + k/3)}$. Thus, $\theta_0$ is of order 
$(t_i/t_0)^{(5/6 + k/3)}$, or, in terms of $r(t_i)$,

\begin{equation}
\theta_0 \sim (r(t_i)/M_P)^{(\frac{5 + 2k}{6 + 3k})}.
\end{equation}

\noindent
As was seen in section 2, the factor $\theta_0^2$ tells how
much the energy of the coherent axion oscillations was
suppressed by the time the axion potential fully turned on at $t_0$.
We will call this suppression factor $S_{before}$. 

\begin{equation}
S_{before} \approx \theta_0^2 \sim 
(r(t_i)/M_P)^{\frac{2}{3}(\frac{5 + 2k}{2 + k})}.
\end{equation} 
  
\noindent
If this were the only suppression of the axion energy, solving the
axion energy problem would require that
$\theta_0 \stackrel{_<}{_\sim} 10^{-7/2}$. With $k = 5/2$
this gives $r(t_i) \stackrel{_<}{_\sim} 2 \times 10^{14}$ GeV.
That is, the initial value of ``$f_a$" can have been quite near
the grand unification scale. By $t_0$ this would have increased,
according to Eq. 12, to a value $r(t_0) \sim \mu_0^2 t_0 \theta_0 \sim
\theta_0 M_P \sim 3 \times 10^{15}$ GeV. One possibility, which we will
call Case I, is that
the radial mode stopped evolving at that point, because its potential
has a minimum there. Another possibility, which we will call Case II, 
is that $r$ continued to increase to some final value near the Planck 
scale. That would mean that the axion energy would have been further 
suppressed by the evolution of $r$ in the period $t > t_0$. Since
$r(t_0) \sim \theta_0 M_P$, and it is assumed that $r(t_f) \sim M_P$,
there is an increase of $r$ by a factor of $\theta_0^{-1}$ in this
period. It is easily shown that the energy of the axion field in
a comoving volume varies inversely with $r$ in this period (as was seen
already in section 3), so that the further suppression of the
axion energy, which we shall call $S_{after}$, is given by

\begin{equation}
S_{after} \approx \theta_0,
\end{equation}

\noindent
or

\begin{equation}
S_{total} = S_{before} S_{after} \approx \theta_0^3.
\end{equation}

\noindent
In Case II, therefore, it is only necessary that $\theta_0 
\stackrel{_<}{_\sim} 10^{-7/3}$, meaning that
$r(t_i) \stackrel{_<}{_\sim} 10^{16}$ GeV. Case I and Case II are,
in a sense, the extreme cases. One can consider intermediate cases
as well. But one sees that in general the relaxing axion scenario 
would have $f_a$
starting out in the range $10^{14}$ to $10^{16}$ GeV, near the grand
unification scale, and evolving to higher values.

\section{The flatness of the radial potential}

The mechanism described in the preceding sections depends 
crucially on the assumption that the potential in the radial
direction is nearly flat. For the mechanism to work, the
centrifugal term in the equation of motion for $r$ had to have 
dominated over the force coming from the potential energy of $r$. 
That is, $r \dot{\theta}^2 > \left| V'(r) \right|$.
One can write this as $\rho_{axion} > \left| r V'(r) \right|$,
where $\rho_{axion}$ is the energy in the coherent axion oscillations.

At some point this condition was no longer satisfied and the
radial field's evolution was controlled by $V(r)$.
Unless the radial field was at that point overdamped (not so 
in the cases of interest) it would have started to oscillate about
the minimum of $V(r)$. For reasonable potentials (where one assumes
that the cosmological constant problem has somehow been solved) one would
expect that $V(r) \sim \left| r V'(r) \right|$, and therefore the 
energy in the
coherent oscillations of the radial mode were also of that order.
Consequently, when the coherent radial oscillations began, the energy 
in them was typically of the same order as the energy in the coherent 
axion oscillations. Thus the coherent radial oscillations do not in
themselves pose a cosmological problem. 

However, for the mechanism to work at all it is necessary that
the centrifugal term in the equation of motion of $r$
dominated over the potential term for a sufficiently long time to 
solve the axion energy problem. This puts a constraint on the 
flatness of $V(r)$.

In Case I, it is assumed that the centrifugal term drove $r$ 
until $t_0$, when the temperature was of order $\mu_0$. At that
time $\rho_{axion}$ had to have been less than about $10^2 \rho_B \sim
10^{-8} \mu_0^3 m_p \sim 10^{-10} {\rm GeV}^4$. Thus,
at $t_0$ it must also have been that $\left|
r V'(r) \right| \stackrel{_<}{_\sim}
10^{-10} {\rm GeV}^4$.

In Case II, the centrifugal term is assumed to have dominated until
$r$ got to be of order $M_P$. Since $r(t_0) \sim \theta_0 M_P$,
and $r$ grew as $t^{1/6}$ for $t > t_0$, this happened at
a time $t_f \sim \theta_0^{-6} t_0$. Assuming a radiation
dominated universe, $T(t_f) \sim \theta_0^3 T(t_0) \sim
S_{total} \mu_0 \sim 10^{-7} \mu_0$. Therefore, at $t_f$ it must 
have been that $\left| r V'(r) \right| \stackrel{_<}{_\sim}
10^{-8} T(t_f)^3 m_p \sim 10^{-31} {\rm GeV}^4$.

How can the potential be that flat? Certainly it is trivial to
arrange that the radial direction be flat in the supersymmetric 
limit. The real problem is to insulate the radial mode from
supersymmetry breaking. This is easiest to do if
supersymmetry is broken at low energies, as it is in models with
gauge-mediated supersymmetry breaking.$^6$ In such a model, the
ordinary quarks must be split from their supersymmetry partners
by an amount that is of order $10^2$ to $10^3$ GeV, which scale we 
will call $m_0$. Since the axion sector must couple directly
or indirectly to the quark sector for the Peccei-Quinn symmetry
to have a QCD anomaly, supersymmetry breaking will be fed into the radial
mode of the axion sector through quark loops. Typically, then, the
radial mode of the axion sector will acquire a potential
of the form $\epsilon m_0^4 \ln (r/m_0)$. $\epsilon$
is model-dependent and is smaller the more indirectly and
the more weakly the axion sector couples to the quark sector.
In Case I, one has that $\epsilon m_0^4 \stackrel{_<}{_\sim}
10^{-10} {\rm GeV}^4$, implying (if $m_0 \sim 1$ TeV) that 
$\epsilon \stackrel{_<}{_\sim} 10^{-22}$. In Case II, one 
has the more severe constraint that $\epsilon \stackrel{_<}{_\sim} 
10^{-43}$. 

To see how such small values of $\epsilon$ might be achieved, 
consider first a
conventional axion model where the radial mode has a tree-level
potential. The relevant terms in the superpotential would
have the following general structure:

\begin{equation}
W_{axion} = g S \overline{Q} Q + W_S,
\end{equation}

\noindent
where $Q$ and $\overline{Q}$ are lefthanded quark and anti-quark
superfields, and $S$ is a superfield containing the axion.
$W_S$ is some set of terms, generally involving other fields,
which has the effect of fixing $\left| \langle S \rangle \right|$
to have some value $M$. The phase of $S$ is, however, assumed not
to be fixed except by QCD instanton effects. One can write
$S = (M + \tilde{S}) e^{i \theta}$, where here we mean by $S$ the
bosonic component, and the axion field is $M \theta$. The radial
mode $\tilde{S}$ would typically have some mass of order $M$.

Consider, now, a somewhat different model, with the corresponding
terms in the superpotential being the following

\begin{equation}
W_{axion}  = g S \overline{Q} Q + W_S + g' (S A - M' B) Y,
\end{equation}

\noindent
where $S$, $A$, $B$, and $Y$ are all gauge singlets, and as before
$W_S$ has the effect of making $\left| \langle S 
\rangle \right| = M$
but leaving the phase of $S$ undetermined. 
Suppose that both $M$ and $M'$ are very large compared to
the scale of supersymmetry breaking. It is apparent that,
since $S$ has a Peccei-Quinn charge, so must either $A$ or $B$.
Therefore, if $\langle A \rangle$ and $\langle B \rangle$ are 
larger than $M$, they rather than $\langle S \rangle$ control
the value of $f_a$. 

Assume that $Y$ has 
no other couplings in the superpotential. Then one of the terms in
the scalar potential is $\left| g' (S A - M' B) \right|^2$. 
This term fixes only the ratio of $A$ and $B$ and leaves them
otherwise undetermined. There is, therefore, a direction that is
flat in the supersymmetric limit along which $A$ and $B$ can 
become arbitrarily large. This flat direction would play
the role of $r$ in this model. 

Writing $S = (M + \tilde{S}) e^{i \theta}$, $A = r e^{i \alpha}$, 
and $B = (\frac{M}{M'} r + \tilde{B}) e^{i (\alpha + \theta + \delta)}$,
the aforementioned term can easily be found to give a mass to
the fields $\tilde{S}$, $\tilde{B}$, and $\tilde{\delta} \equiv r \delta$. 
In particular, one finds that $\left| g' (S A - M' B) \right|^2
= g^{'2} M^2 \tilde{\delta}^2 + g^{'2} (\langle r \rangle \tilde{S} -
M' \tilde{B})^2 +$ terms of cubic and higher order. In addition, there
is the mass term for $\tilde{S}$ coming from $W_S$.
The phase $\alpha$ is an exact goldstone mode,
while $\theta$, since it corresponds to an anomalous $U(1)$ by
virtue of its coupling to the quarks, is the axion mode. The
scalar field $r$ has (before supersymmetry breaking) a flat
potential. It is easily seen that the Lagrangian terms for
$r$ and $\theta$ have (after suitable rescaling of fields)
essentially the same form as those shown
in Eq. 1. Thus this model is an implementation of the 
relaxing axion mechanism. The question is how large a potential
$r$ gets in this model. If supersymmetry breaking is mediated from
some hidden sector by Standard Model gauge interactions, then
only $\overline{Q}$ and $Q$ will directly feel it. The splittings
in the supermultiplet $S$ will only arise through a quark loop, and the 
splittings in the multiplets $A$ and $B$ will only arise through
diagrams involving both an $S$ loop and a quark loop.
One would expect that $\epsilon \sim (g^2/16 \pi^2)(g^{'2}/16 \pi^2)$.
This can easily be as small as required. For example, if $M$ and
$M'$ are near the unification scale, the coupling $g$ could even
be as small as $10^{-14}$. 

It seems to be possible to shield the radial mode from supersymmtry 
breaking even more thoroughly. As an extreme case one could imagine a
whole series of gauge-singlet sectors separating the quark sector 
from the axion sector: $S_{quark} \longleftrightarrow S_1 
\longleftrightarrow ...
\longleftrightarrow S_n \longleftrightarrow S_{axion}$, where
the $S$'s denote various sectors, and the arrows represent a coupling
in the superpotential between two sectors. In this case, supersymmetry 
breaking in the radial mode would be an $n$-loop effect. On the other 
hand, because of the couplings between sectors, there are fields in each
sector that have non-trivial Peccei-Quinn charges. Thus the axion
field, though removed by several steps from the quarks, will
nonetheless get a potential from QCD instanton effects that goes as
$- \mu_0^4 \cos (N \theta)$, where $N$ is an integer.

From these examples, it seems that there is no reason in principle
why the radial mode could not be sufficiently flat to allow
the relaxing axion mechanism to work. It is an interesting question
whether this flat radial direction can be identified with other
flat directions that have been discussed in recent years.
Could it be, to mention two obvious examples, the dilaton of 
superstring theory or the ``quintessence" field? In any event, given 
the increasing role that very flat potentials have played in 
particle physics and cosmology (inflatons, quintessence, dilatons,
moduli, and the axion itself), it is suggestive that a flat
radial direction can allow the breaking of the Peccei-Quinn
symmetry to take place at the unification scale or even higher.

\vspace{1cm}

This work was supported in part by the Department of Energy
under contract No. DE-FG02-91ER-40626.

\section*{References}

\begin{enumerate}

\item  S. Weinberg, {\it Phys. Rev. Lett.} {\bf 40}, 223 (1978);
F. Wilczek, {\it ibid.}, 271. For a review see ``The Strong CP
Problem", by R.D. Peccei, in {\it CP Violation}, ed. C. Jarlskog
(World Scientific, 1989), pp. 503-551.

\item  J. Preskill, M. Wise, and F. Wilczek, {\it Phys. Lett.}
{\bf 120B}, 127 (1983); L. Abbott and P. Sikivie, {\it ibid.}, p. 133;
M. Dine and W. Fischler, {\it ibid.}, p. 137.

\item  P. Steinhardt and M. Turner, {\it Phys. Lett.} {\bf 129B}, 51 (1983);
G. Lazarides, R.K. Schaefer, D. Seckel, and Q. Shafi, {\it Nucl. Phys.} 
{\bf B346}, 193 (1990); S.-Y. Pi, {\it Phys. Rev.} {\bf D24}, 1725 (1984);
A. Linde, {\it Phys. Lett.} {\bf 201B}, 437 (1988);
G. Dvali, IFUP-TH 21/95 (unpublished); S.M. Barr, K.S. Babu, and
D. Seckel, {\it Phys. Lett.} {\bf 336B} 213 (1994).

\item M. Bronstein, {\it Physikalische Zeitschrift Sowjet Union}
{\bf 3}, 73 (1933); M. \"{O}zer and M.O. Taha, {\it Nucl. Phys.}
{\bf B287}, 797 (1987); B. Ratra and P.J.E. Peebles, {\it Phys.
Rev.} {\bf D37}, 3406 (1988); J.A. Freiman, C.T. Hill, and
R. watkins, {\it Phys. Rev.} {\bf D46}, 1226 (1992); 
R. Caldwell, R. Dave, and P.J. Steinhardt, {\it Phys. Rev. Lett.}
{\bf 80}, 1582 (1998).

\item D.J. Gross, R.D. Pisarski, and L.G. Yaffe, {\it Rev.
Mod. Phys.} {\bf 53}, 43 (1981). 

\item  For a review see A. Nelson, {\it Nucl. Phys. B (proc. Suppl.}
62A-C, 261 (1998).

\end{enumerate}

\end{document}